\documentclass[preprint,showpacs,preprintnumbers,amsmath,amssymb]{revtex4}
\usepackage{graphicx}
\usepackage{dcolumn}
\usepackage{bm}
\usepackage{mathrsfs}
\usepackage{amsmath,amssymb,graphics,epsfig,subfigure}

\begin{document}

\title{Geometric curvatures of plane symmetry black hole}

\author{Shao-Wen Wei\footnote{E-mail: weishaow06@lzu.cn} ,
        Yu-Xiao Liu\footnote{Corresponding author. E-mail:
                             liuyx@lzu.edu.cn},
        Chun-E Fu\footnote{E-mail: fuche08@lzu.cn}
        and
        Hai-Tao Li\footnote{E-mail: liht07@lzu.cn}                             }
\affiliation{
    Institute of Theoretical Physics, Lanzhou University,
           Lanzhou 730000, P. R. China}

\begin{abstract}
In this paper, we study the properties and thermodynamic stability
of the plane symmetry black hole from the viewpoint of geometry.
Weinhold metric and Ruppeiner metric are obtained, respectively. The
Weinhold curvature gives phase transition points, which correspond
to the first-order phase transition only at $N=1$, where $N$ is a
parameter in the plane symmetry black hole. While the Ruppeiner one
shows first-order phase transition points for arbitrary $N\neq 1$.
Both of which give no any information about the second-order phase
transition. Considering the Legendre invariant proposed by Quevedo
et. al., we obtain a unified geometry metric, which gives a
correctly the behavior of the thermodynamic interactions and phase
transitions. The geometry is also found to be curved and the scalar
curvature goes to negative infinity at the Davies' phase transition
points when the logarithmic correction is included.
\end{abstract}


\maketitle

\section{Introduction}
\label{secIntroduction}

Several decades ago, the original works of Bekenstein and Hawking
showed that the black hole is indeed a thermodynamics system
\cite{Bekenstein1973prd,Hawking1975cmp}. It was also found that the
black hole satisfies four laws of the elementary thermodynamics with
regarding the surface gravity and the outer horizon area as its
temperature and entropy, respectively \cite{Hawking1973cmp}.
Although, it is widely believed that a black hole is a thermodynamic
system, the statistical origin of the black hole entropy is still
one of the most fascinating and controversial subjects today.

The investigation of thermodynamic properties of black holes is also
a fascinating subject. Much of work had been carried out on the
stability and phase transitions of black holes. It is generally
thought that the local stability of a black hole is mainly
determined by its heat capacity. Negative heat capacity usually
gives a thermodynamically unstable system and the positive one
implies a local stable one. The points, where the heat capacity
diverges, are usually consistent with the Davies' points, where the
second-order phase transition takes place \cite{Davies1977}.

The properties of a thermodynamic system can also be studied with
the ideas of geometry. Weinhold \cite{Weinhold1975jcp} firstly
introduced the geometrical concept into the thermodynamics. He
suggested that a Riemannian metric can be defined as the second
derivatives of internal energy $U$ with respect to the entropy and
other extensive quantities of a thermodynamic system. However, it
seems that the Weinhold geometry has not much physical meanings. Few
years later, Ruppeiner \cite{Ruppeiner1979pra} introduced another
metric, which is analogous to the Weinhold one. The thermodynamic
potential of the Ruppeiner geometry is the entropy $S$ of the
thermodynamic system rather than the internal energy $U$. In fact,
the two metrics are conformally related to each other
\begin{equation}
 ds^{2}_{R}=\frac{1}{T}ds^{2}_{W},
\end{equation}
with the temperature $T$ as the conformal factor. The Ruppeiner
geometry had been used to study the ideal gas and the van der Waals
gas. It was shown that the curvature vanishes for the ideal gas
whereas, for the van der Waals gas, the curvature is non-zero and
diverges only at those points, the phase transitions take place(for
details see the review paper \cite{Ruppeiner1995rmp}). The black
hole, as a thermodynamics system, has been extensively investigated.
The Weinhold geometry and the Ruppeiner geometry were obtained for
various black holes and black branes
\cite{Davies1977,Weinhold1975jcp,Ruppeiner1979pra,Ruppeiner1995rmp,
Ferrara1997npb,cai1999prd,Aman2003grg,Johnston2003app,
Arcioni2005prd,Shen2007ijmpa,Aman2006grg,Aman2006prd,
Mirza2006jhep,Aman2008eps,Aman2008,Medved2008mpla,
Myung2008plb,Gergely2008,Wei2009prd,Biswas2009,Sarkar2008jhep,Bellucci2008,
Aman2007jpcs,Ruppeiner2008prd,Vacaru1999}. In particular, it was
found that the Ruppeiner geometry carries the information of phase
structure of a thermodynamic system. In general, its curvature is
singular at the points, where the phase transition takes place.
However, for the rotating Banados-Teitelboim-Zanelli (BTZ) and
Reissner-Nordstr$\ddot{o}$m (RN) black holes, the cases are quite
different. The Ruppeiner geometry gives a vanished curvature, which
means there exists no thermodynamic interactions and no phase
transition points. But the two kinds of black hole  do exist the
phase transition points. For this contradiction, many researches has
been carried out to explain it. The main focus is on the
thermodynamic potential, which is generally believed to be the
internal energy $U$ rather than the mass $M$ for simple. For the
Reissner-Nordstr$\ddot{o}$m black hole, it was argued in
\cite{Mirza2006jhep} that, the thermodynamic curvature should be
reproduced from the Kerr-Newmann anti-de sitter black hole with the
angular momentum $J\rightarrow 0$ and cosmological constant
$\Lambda\rightarrow 0$. Another explanation of the contradiction was
presented by Queved et. al. few years ago
\cite{Quevedo2007jmp,Quevedo2008prc}. They pointed out that the
origin of the contradiction is that the Weinhold metric and
Ruppeiner metric are not Legendre invariant. A Legendre invariant
metric was introduced by them, which could reproduce correctly the
behavior of the thermodynamic interactions and phase transitions for
the BTZ and RN black holes \cite{Quevedo2009prd,Quevedo2008grg} and
other black hole configurations and models \cite{Alvarez2008prd,
Quevedo2008jhep,Quevedo2008,2Quevedo2009prd}.

Another interesting and important question of this field is how the
geometry behaves when including the corrected term of the entropy.
It is generally believed that, for a canonical ensemble, there
exists a logarithmic corrected term to the entropy \cite{Huang1963}.
Considering the correction term, the geometry structure was studied
in \cite{Quevedo2009prd,Sarkar2006jhep} for the BTZ black hole.
Especially, its Ruppeiner curvature will become non-zero when the
logarithmic correction is included. The aim of this paper is to
study the phase transitions and geometry structure of the plane
symmetry black hole. Firstly, we study the thermodynamic stability
of the plane symmetric black hole. It is shown that there always
exist locally thermodynamically stable phases and unstable phases
for the plane symmetric black hole due to suitable parameter
regimes. Then, three different geometry structures are obtained. The
Weinhold curvature gives phase transition points, which correspond
to that of the first-order phase transition only at $N=1$. While the
Ruppeiner one shows first-order phase transition points for
arbitrary $N\neq 1$. Both of which give no any information about the
second-order phase transition. Considering the Legendre invariant,
we obtain a unified geometry metric, which gives a correct behavior
of the thermodynamic interactions and phase transitions. It is found
that the curvature constructed from the unified metric goes to
negative infinity at the Davies' points, where the second-order
phase transition takes place. The geometry structure is also studied
as the logarithmic correction is included. The result shows that the
logarithmic correction term has no affect on the unified geometry
depicts the phase transitions of the plane symmetric black hole. We
also show that the absolute values of the charge $Q$ at the
divergence points of the curvature $\mathcal{R}'_{L}$ will decrease
with the increase of the entropy corrected parameter $\gamma$ for
large $S$.

The paper is organized as follows. In Sec. \ref{Thermodynamic}, we
first review some thermodynamic quantities of the plane symmetric
black hole. The thermodynamic stability is also studied. In Sec.
\ref{geometry}, both the Weinhold and Ruppeiner geometry structures
are obtained. However, they fail to give the information about the
second-order phase transition points. For the reason, we give a
detail analysis and obtain a new Legendre invariant metric structure
which could give a well description of the thermodynamic
interactions and phase transitions in Sec.
\ref{Geometrothermodynamics}. Including the logarithmic corrected
term, the geometry structure is considered in Sec.
\ref{fluctuations}. Finally, the paper ends with a brief conclusion.

\section{Thermodynamic Quantities and Thermodynamic Stability of the Plane Symmetric Black Hole }
\label{Thermodynamic}

In this section, we will present the thermodynamic quantities and
other properties of the plane symmetric black hole. The
thermodynamic stability of it is also discussed. The action
depicting the plane symmetric black hole is given by
\begin{equation}
 S=\frac{1}{16\pi}\int d^{4}x \sqrt{-g}\big(\mathcal{R}-2(\nabla \varphi)^{2}
   -2\Lambda e^{2b\varphi}-e^{-2a\varphi}F^{2}\big).
\end{equation}
where $\varphi$ is a dilaton field and $a$, $b$ are constants. The
negative cosmological constant $\Lambda=-3\alpha^{2}$. Static plane
symmetric black hole solutions in this theory were first given in
\cite{Cai1996prd} (some detail works for this black hole can also be
found in
\cite{Miranda2008jhep,Zeng2009ctp,Zeng2009IJTP,Lemos2004prd})
\begin{equation}
 ds^{2}=-f(r)dt^{2}+f^{-1}(r)dr^{2}+g(r)(dx^{2}+dy^{2}).
\end{equation}
The metric functions are given by, respectively,
\begin{eqnarray}
 f(r)&=&-\frac{4\pi M}{N \alpha^{N}}r^{1-N}
      +\frac{6\alpha^{2}}{N(2N-1)}r^{N}
      +\frac{2Q^{2}}{N\alpha^{2N}}r^{-N},\\
 g(r)&=&(r\alpha)^{N}.
\end{eqnarray}
where $N\in (\frac{1}{2},2)$. The parameters $M$ and $Q$ are the
mass and charge of the black hole. The event horizon is located at
$f(r_{h})=0$ and the radius $r_{h}$ satisfies
\begin{equation}
 \frac{3\alpha^{2}}{2N-1}r_{h}^{2N}
      -\frac{2\pi
      M}{\alpha^{N}}r_{h}+\frac{Q^{2}}{\alpha^{2N}}=0.\label{event}
\end{equation}
In general, there exist two event horizons, the inner event horizon
and the outer event horizon. Under the extreme case, the two horizon
will merge into each other. Here, we have denoted $r_{h}$ as the
radius of outer event horizon.

\begin{figure*}[htb]
\begin{center}
\includegraphics[width=5.6cm,height=4cm]{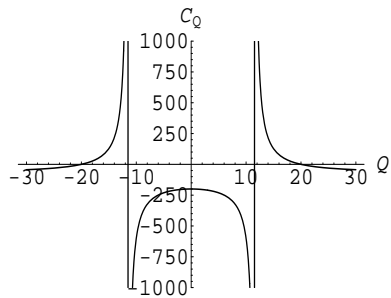}
\end{center}
\vskip -4mm \caption{The behavior of the heat capacity $C_{Q}$, with
$N=1$, $\alpha=\frac{\sqrt{3}}{3}$ and $S=100$. $C_{Q}$ is singular
at $Q=\pm 11.5470$ and vanishes at $Q=\pm 20$.} \label{figcapacity}
\end{figure*}

The surface area of the event horizon corresponds to unit $xoy$
plane \cite{Zeng2009IJTP}
\begin{equation}
\mathcal{A}=(\alpha r_{h})^{2N}.
\end{equation}
From Eq. (\ref{event}), the mass can be expressed in the form
\begin{equation}
 M=\frac{3 \alpha^{N+2}}{2\pi (2N-1)}r_{h}^{2N-1}
 +\frac{Q^{2}}{2\pi \alpha^{N}}r_{h}^{-1}.\label{mass}
\end{equation}
With the relation between area and entropy, i.e.
$S=\frac{\mathcal{A}}{4}$, we can obtain
\begin{equation}
 r_{h}=\frac{1}{\alpha}(4S)^{\frac{1}{2N}}.\label{radius}
\end{equation}
Substituting Eq. (\ref{radius}) into (\ref{mass}), the mass can be
obtained as a function of entropy $S$ and charge $Q$ in the form
\begin{equation}
 M=\frac{3 \alpha^{3-N}}{2\pi (2N-1)}(4S)^{\frac{2N-1}{2N}}
 +\frac{Q^{2}}{2\pi
 \alpha^{N-1}}(4S)^{-\frac{1}{2N}}.\label{lastmass}
\end{equation}
From the energy conservation law of the black hole
\begin{equation}
 dM=TdS+\phi dQ,
\end{equation}
the relevant thermodynamic variables,  the temperature and electric
potential are obtained
\begin{eqnarray}
 T&=&\left(\frac{\partial M}{\partial S}\right)_{Q}
   =\frac{(12\alpha^{2}S-Q^{2})\alpha^{1-N}}
   {2^{2+\frac{1}{N}}\pi NS^{1+\frac{1}{2N}}},\label{temperature}\\
 \phi&=&\left(\frac{\partial M}{\partial Q}\right)_{S}
      =\frac{\alpha^{1-N}Q}{2^{\frac{1}{N}}\pi S^{\frac{1}{2N}}}.
\end{eqnarray}
For a given charge $Q$, the heat capacity has the expression
\begin{equation}
 C_{Q}=-\frac{2NS(12\alpha^{2}S-Q^{2})}{12S\alpha^{2}-(1+2N)Q^{2}}\;,\label{2heatcapacity}
\end{equation}
with the zero-points and singular points
\begin{eqnarray}
 Q^{2}&=&12\alpha^{2}S,\quad\quad \mbox{(zero-points)}\label{zeropoint}\\
 Q^{2}&=&\frac{12\alpha^{2}S}{2N+1},\quad \;\mbox{(singular points)}\label{singularpoint}
\end{eqnarray}
respectively. The heat capacity $C_{Q}$ vanishes at
$Q^{2}=12\alpha^{2}S$, which is considered to be the first-order
phase transition points. It is generally believed that the Davies'
points where the second-order phase transition takes place
correspond to the diverge points of heat capacity. So the heat
capacity (\ref{2heatcapacity}) may indicate that the second-order
phase transition takes place at $Q^{2}=\frac{12\alpha^{2}S}{2N+1}$.
The heat capacity also contains the information of the local
stability of the black hole thermodynamics. The negative heat
capacity always implies an unstable thermodynamics system and the
positive one shows a stable system. Here, we would like to give a
brief discussion about the local stability of the plane symmetric
black hole. For $Q^{2}>12\alpha^{2}S$, the numerator of the heat
capacity (\ref{2heatcapacity}) is negative, while the denominator is
positive, which gives a negative heat capacity. For
$Q^{2}<\frac{12\alpha^{2}S}{2N+1}$, the numerator is positive, but
the denominator turns to negative, which also shows a negative heat
capacity. So, in both cases, the heat capacity $C_{Q}$ implies an
unstable black hole thermodynamics. When $|Q|\in
(2\sqrt{3}\alpha\sqrt{\frac{S}{2N+1}},2\sqrt{3}\alpha\sqrt{S})$,
both the numerator and denominator are positive, which implies a
stable black hole thermodynamics. The behavior of the heat capacity
$C_{Q}$ can be directly found from Fig. \ref{figcapacity}. For
Larger and smaller values of $|Q|$, the heat capacity $C_{Q}$ is
negative. While in the middle zone, it is positive. In summary, we
have found that there always exist locally thermodynamically stable
phases and unstable phases for the plane symmetric black hole due to
suitable parameter regimes.

\section{Weinhold geometry and Ruppeiner geometry of the Plane Symmetric Black Hole}
\label{geometry}

In this section, we would like to study the Weinhold geometry and
Ruppeiner geometry of the plane symmetric black hole. In the first
step, we will show the Weinhold geometry. Then, using the conformal
relation, we could obtain the Ruppeiner geometry. The Weinhold
geometry is charactered by the metric
\begin{equation}
 ds_{W}^{2}=\frac{\partial^{2} M}{\partial S^{2}}dS^{2}
            +2\frac{\partial^{2} M}{\partial S \partial Q}dSdQ
            +\frac{\partial^{2} M}{\partial Q^{2}}dQ^{2},
\end{equation}
where the index $W$ denotes the Weinhold geometry. Here, we have
made the choice that the mass $M$ corresponds to the thermodynamic
potential and with the extensive variables entropy $S$ and charge
$Q$.

\begin{figure*}[htb]
\begin{center}
\includegraphics[width=5.6cm,height=4cm]{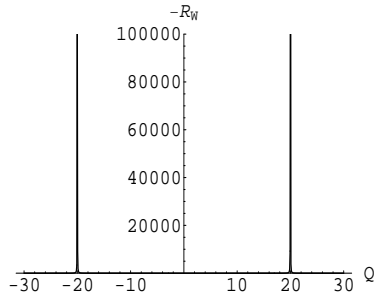}
\end{center}
\vskip -4mm \caption{The negative Weinhold curvature
$\mathcal{R}_{w}$ vs. the charge Q with $N=1$,
$\alpha=\frac{\sqrt{3}}{3}$ and $S=100$. The divergence points are
at $Q=\pm 20$.} \label{RW}
\end{figure*}

Using Eq. (\ref{lastmass}), the Weinhold metric can be obtained in
the form
\begin{eqnarray}
g_{W}=-\frac{\alpha^{1-N}}{2^{\frac{1}{N}}S^{\frac{1}{2N}}}
\left(
  \begin{array}{cc}
    \frac{12\alpha^{2}S-(2N+1)Q^{2}}{8\pi N^{2}S^{2}} & \frac{Q}{2\pi N S} \\
    \frac{Q}{2\pi N S} & -\frac{1}{\pi} \\
  \end{array}
\right).
\end{eqnarray}
Its determinant is $\det(g)=-\frac{\alpha^{2-2N}[12\alpha^{2}
S-(2N-1)Q^{2}]}{2^{3+\frac{2}{N}}S^{2+\frac{1}{N}}\pi^{2}N^{2}}$.
It can be seen that the determinant disappears as the heat
capacity vanishes only at $N=1$. A simple calculation shows that
the Christoffel symbols are
\begin{eqnarray}
 \Gamma^{S}_{SS}&=&-\frac{(2N+1)\big[12\alpha^{2}S-(4N-1)Q^{2}\big]}
                         {4NS\big[12\alpha^{2}S-(2N-1)Q^{2}\big]},\quad
 \Gamma^{S}_{QS}=\Gamma^{S}_{SQ}=-\frac{2NQ}{12\alpha^{2}S-(2N-1)Q^{2}},\nonumber\\
 \Gamma^{S}_{QQ}&=&-\frac{2NS}{12\alpha^{2}S-(2N-1)Q^{2}},\quad\quad\quad\quad\quad\quad
 \Gamma^{Q}_{SS}=-\frac{(2N+1)Q^{3}}{4NS^{2}\big[12\alpha^{2}S-(2N-1)Q^{2}\big]},\quad\\
 \Gamma^{Q}_{SQ}&=&\Gamma^{Q}_{QS}
       =-\frac{12\alpha^{2}S+(2N+1)Q^{2}}{4S\big[12N\alpha^{2}S-N(2N-1)Q^{2}\big]},\quad
 \Gamma^{Q}_{QQ}=\frac{Q}{12\alpha^{2}S-(2N-1)Q^{2}},\quad \nonumber
\end{eqnarray}
where the Christoffel symbols is calculated with
\begin{eqnarray}
 \Gamma^{\lambda}_{\mu\nu}&=&\frac{1}{2}g^{\lambda\tau}(g_{\nu\tau,\mu}+g_{\mu\tau,\nu}-g_{\mu\nu,\tau}).
\end{eqnarray}
The Riemannian curvature tensor, Ricc curvature and scalar curvature
are given, respectively
\begin{eqnarray}
 &&R^{\mu}_{\sigma\nu\tau}=\Gamma^{\mu}_{\sigma\nu,\tau}-\Gamma^{\mu}_{\sigma\tau,\nu}
                         +\Gamma^{\mu}_{\lambda,\tau}\Gamma^{\lambda}_{\sigma,\nu}
                         -\Gamma^{\mu}_{\lambda,\nu}\Gamma^{\lambda}_{\sigma,\tau},\nonumber\\
 &&R_{\mu\nu}=R^{\lambda}_{\mu\lambda\nu},\label{expression}\\
 &&R=g^{\mu\nu}R_{\mu\nu}.\nonumber
\end{eqnarray}
With Eq. (\ref{expression}), we get the scalar curvature
\begin{equation}
 \mathcal{R}_{W}=-\frac{24\; 2^{\frac{1}{N}}N\pi S^{1+\frac{1}{2N}}\alpha^{1+N}}
                {\big[12\alpha^{2}S-(2N-1)Q^{2}\big]^{2}}.
\end{equation}
This curvature is always negative for any values of charge $Q$ and
positive entropy $S$. It diverges at
$Q^{2}=\frac{12\alpha^{2}S}{2N+1}$, which consists with the
first-order transition points (\ref{zeropoint}) reproduced from the
capacity $C_{Q}$ only at $N=1$. Its behavior can be seen in Fig.
\ref{RW}. However, it implies no information about the second-order
phase transition. So, it is natural to ask how the Ruppeiner
curvature behaves. Could it gives the proper phase transition
points?

\begin{figure*}[htb]
\begin{center}
\includegraphics[width=5.6cm,height=4cm]{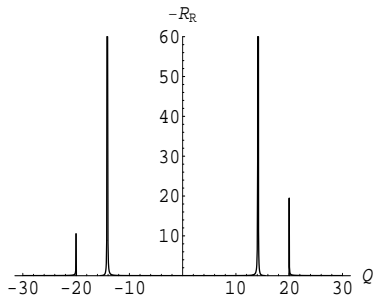}
\end{center}
\vskip -4mm \caption{The negative Ruppeiner curvature
$\mathcal{R}_{R}$ vs. the charge Q with $N=\frac{3}{2}$,
$\alpha=\frac{\sqrt{3}}{3}$ and $S=100$. The divergence points are
at $Q=\pm 20,\pm 14.1421$.} \label{Ruppeiner}
\end{figure*}

With that question, we now turn to the Ruppeiner geometry of the
plane symmetric black hole. Recalling the conformal relation between
the Ruppeiner geometry and the Weinhold geometry, we obtain the
Ruppeiner metric
\begin{eqnarray}
g_{R}=\frac{1}{T}g_{W}=\left(
  \begin{array}{cc}
    -\frac{12\alpha^{2}S-(2N+1)Q^{2}}{2NS(12\alpha^{2}S-Q^{2})} & -\frac{2Q}{12\alpha^{2}S-Q^{2}} \\
    -\frac{2Q}{12\alpha^{2}S-Q^{2}} & -\frac{4NS}{12\alpha^{2}S-Q^{2}} \\
  \end{array}
\right),
\end{eqnarray}
where the index $R$ denotes the Ruppeiner geometry. After some
calculations, we obtain the Ruppeiner curvature
\begin{equation}
 \mathcal{R}_{R}=-\frac{12\alpha^{2}Q^{2}(N-1)\big[36S\alpha^{2}-(4N-1)Q^{2}\big]}
                 {(12\alpha^{2}S-Q^{2})\big[12S\alpha^{2}-(2N-1)Q^{2}\big]^{2}}.
\end{equation}
It is obvious that the curvature will be zero at $N=1$. The vanished
thermodynamic curvature $\mathcal{R}_{R}$ implies that there exists
no phase transition points and no thermodynamic interactions appear.
So, the Ruppeiner curvature is not proper to describe the phase
transitions of the plane symmetric black hole at $N=1$. The
divergence of the Ruppeiner curvature is at
$Q^{2}=\frac{12\alpha^{2}S}{2N-1}$ and $Q^{2}=12\alpha^{2}S$, which
can be seen from the Fig. (\ref{Ruppeiner}). The points
$Q^{2}=12\alpha^{2}S$ consist with the zero-points (\ref{zeropoint})
of heat capacity $C_{Q}$. This means that the Ruppeiner curvature
always implies the first-order phase transition points. Like the
Weinhold curvature, the Ruppeiner curvature also implies no any
information about the second-order phase transition.

\section{Unified geometry of the Plane Symmetric Black Hole}
\label{Geometrothermodynamics}

In the previous section, we show that the Weinhold curvature implies
the first-order phase transition points only at $N=1$, while the
Ruppeiner curvature implies the first-order phase transition points
expect $N=1$. Both of the geometry structures fail to give the
second-order phase transition points of the plane symmetric black
hole. Quevedo et. al. pointed out the failure of the two geometries
to describe the second-order phase transition points is that they
are not Legendre invariant, which makes them inappropriate to
describe the geometry of thermodynamic systems
\cite{Quevedo2008grg}. Considering the Legendre invariant, a unified
geometry was presented in \cite{Alvarez2008prd}, where the metric
structure can give a well description of various types of black hole
thermodynamics. So, in this section, we would like to discuss the
unified geometry of the plane symmetric black hole and we want to
know whether it works.

\begin{figure*}[htb]
\begin{center}
\includegraphics[width=5.6cm,height=4cm]{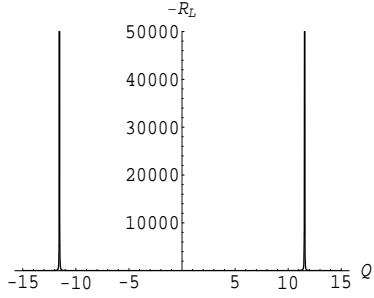}
\end{center}
\vskip -4mm \caption{ The negative unified geometric curvature
$\mathcal{R}_{L}$ vs. the charge Q with $N=1$,
$\alpha=\frac{\sqrt{3}}{3}$ and $S=100$. The divergence points are
at $Q=\pm 11.5470$, which are consistent with the divergence points
of the heat capacity $C_{Q}$. The positive curvature region is at
$|Q|\geq 20$ and it is not shown in this figure.} \label{RL}
\end{figure*}

The unified geometry metric can be expressed as
\begin{eqnarray}
 ds^{2}_{L}&=&\bigg(S\frac{\partial M}{\partial S}
     +Q\frac{\partial M}{\partial Q}\bigg)
     \left(
  \begin{array}{cc}
    -\frac{\partial^{2} M}{\partial S^{2}} & 0 \\
    0 & \frac{\partial^{2} M}{\partial Q^{2}} \\
  \end{array}
\right)
\left(
  \begin{array}{c}
    dS^{2} \\
    dQ^{2} \\
  \end{array}
\right)\nonumber\\
  &=&\frac{\alpha^{2-2N}\big[12\alpha^{2}S-(2N+1)Q^{2}\big]\big[12\alpha^{2}S+(4N-1)Q^{2}\big]}
              {2^{5+\frac{2}{N}}\pi^{2}N^{3}S^{2+\frac{1}{N}}}dS^{2}\nonumber\\
     &&+\frac{\alpha^{2-2N}(12\alpha^{2}S+(4N-1)Q^{2})}{2^{\frac{2(N+1)}
           {N}}\pi^{2}NS^{\frac{1}{N}}}dQ^{2}. \label{unify}
\end{eqnarray}
The index $L$ denotes the curvature reproduced from the Legendre
invariant metric.
This diagonal metric reproduces the thermodynamic curvature
$\mathcal{R}_{L}$, which turns out to be non-zero and the scalar
curvature is
\begin{eqnarray}
 \mathcal{R}_{L}&=&\frac{192\;4^{\frac{1}{N}}\pi^{2}\alpha^{2N}NS^{1+\frac{1}{N}}}
     {\big[12\alpha^{2}S-(2N+1)Q^{2}\big]^{2}
      \big[12\alpha^{2}S+(4N-1)Q^{2}\big]^{3}} \nonumber \\
    && \cdot \bigg\{(4N-1)Q^{2}
                \big[(N(4N^{2}-6N-3)-1)Q^{2} \nonumber \\
    &&~             -12((N-5)N-2)\alpha^{2}S\big]
               +144(N-1)^{2}\alpha^{4}S^{2}\bigg\}.\label{curvature}
\end{eqnarray}
The thermodynamic curvature vanishes at $Q^{2}=12S\alpha^{2}$ when
$N=1$, which is just the points of the first-order phase transition.
It is shown that the diverge points are at
$Q^{2}=\frac{12\alpha^{2}S}{2N+1}$, which implies that there exists
second-order phase transitions at that points. This result exactly
consists with that of the heat capacity (\ref{2heatcapacity}). The
detail behavior of $\mathcal{R}_{L}$ can be found in Fig. \ref{RL},
where the singularities are just the divergence points of the heat
capacity $C_{Q}$. Now, we can see that the thermodynamic curvature
$\mathcal{R}_{L}$ reproduced from the Legendre invariant metric
(\ref{unify}) could give an exact description of the second-order
phase transitions of a thermodynamics system. Beside this, we also
expect that this unified geometry description may give more
information about a thermodynamics system.

\section{Inclusion of logarithmic correction}
\label{fluctuations}

In this section, we will discuss the unified geometry of the plane
symmetric black hole when the logarithmic corrected term is
included. For general, we suppose that the correct-entropy formula
is of the form
\begin{eqnarray}
 S'=S-\gamma \ln S. \label{entropy correct}
\end{eqnarray}
The parameter $\gamma$ is a constant. In fact, the origin of the
logarithmic correction term can be accounted by the uncertainty
principle or the tunneling method.

With Eq. (\ref{entropy correct}), the heat capacity
(\ref{2heatcapacity}) is modified to
\begin{eqnarray}
 C'_{Q}=\frac{2NS(S-\gamma)(S-\gamma\ln S)(12S\alpha^{2}-Q^{2}-12\alpha^{2}\gamma\ln
 S)}{A_{2}Q^{2}-A_{1}},
\end{eqnarray}
with
\begin{eqnarray}
 A_{1}&=&12\alpha^{2}(S-\gamma{\ln}S)
      \big[\big(1+2N\ln{S}\big)\gamma^{2}-2(N+1)S{\gamma}+S^{2} \big], \\
 A_{2}&=&\big[2N(1+\ln S)+1\big]\gamma^{2}-2(3N+1)S\gamma+(2N+1)S^{2}.
\end{eqnarray}
The singular points of the heat capacity are determined by
$A_{2}Q^{2}-A_{1}=0$ and are given by
\begin{eqnarray}
 Q^{2}=\frac{A_1}{A_2}.
\end{eqnarray}
If $\gamma=0$, the singular points of the heat capacity will become
Eq. (\ref{2heatcapacity}). Following the Sec.
\ref{Geometrothermodynamics}, we obtain the curvature
$\mathcal{R}'_{L}$:
\begin{eqnarray}
 \mathcal{R}'_{L}=\frac{h(S,Q,N,\gamma)}{K^{3}(A_{2}Q^{2}-A_{1})^{2}},
\end{eqnarray}
where $K=[4N(S-\gamma\ln
S)+\gamma-S]Q^{2}+12\alpha^{2}(S-\gamma)(S-\gamma\ln S)$, and
$h(S,Q,N,\gamma)$ is a complex function and we do not write it here.
It is found that the divergence points for the heat capacity
$C'_{Q}$ and the curvature $\mathcal{R}'_{L}$ consist with each
other, which means the curvature give proper points, where
second-order phase transitions take place. So, it is easy to
summarized that the logarithmic correction term does not affect the
unified geometry to depict the plane symmetry black hole's phase
transitions.

\begin{figure*}[htb]
\begin{center}
\includegraphics[width=3.5cm,height=4cm]{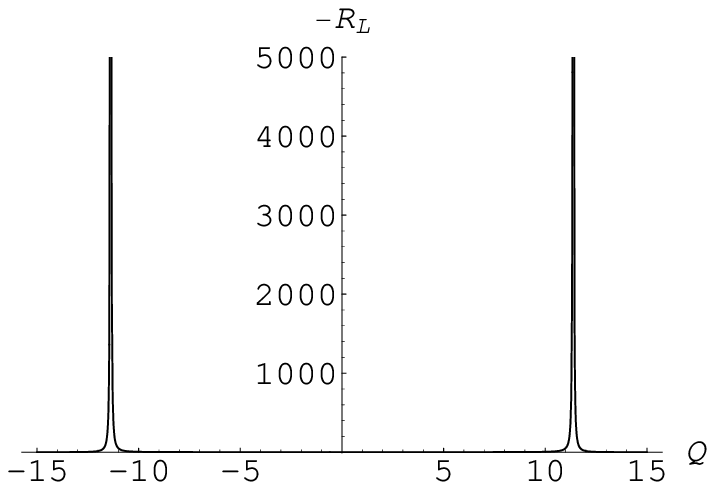}
\includegraphics[width=3.5cm,height=4cm]{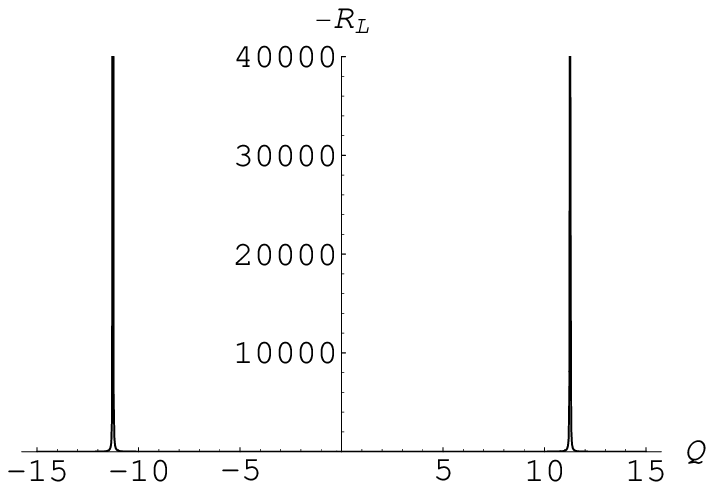}
\includegraphics[width=3.5cm,height=4cm]{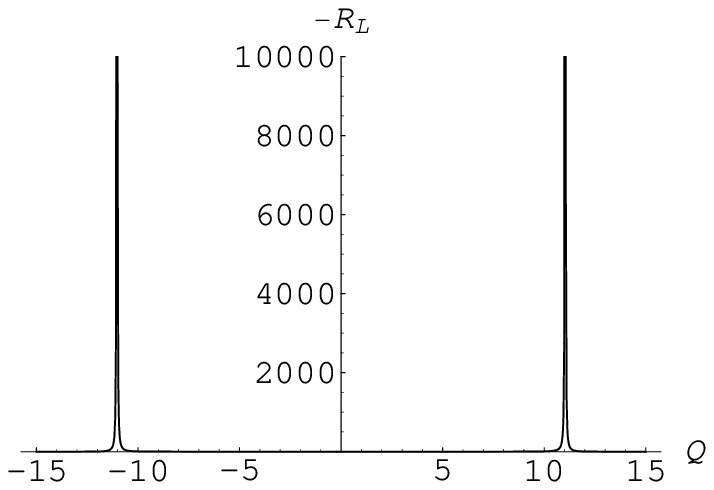}
\end{center}
\vskip -4mm \caption{The negative unified geometric curvature
$\mathcal{R}_{L}$ vs. the charge Q, including the logarithmic
correction. The values of $\alpha$ and the entropy $S$ are the same
in Fig. \ref{RL}. The parameter $\gamma$ is set to
$\frac{1}{2}$(left), $\frac{5}{6}$(middle) and $\frac{3}{2}$(right),
respectively. The divergence points are at $Q=\pm 11.3756$, $\pm
11.2613$ and $\pm 11.0326$. } \label{2RL}
\end{figure*}

\begin{figure*}[htb]
\begin{center}
\includegraphics[width=5.6cm,height=4cm]{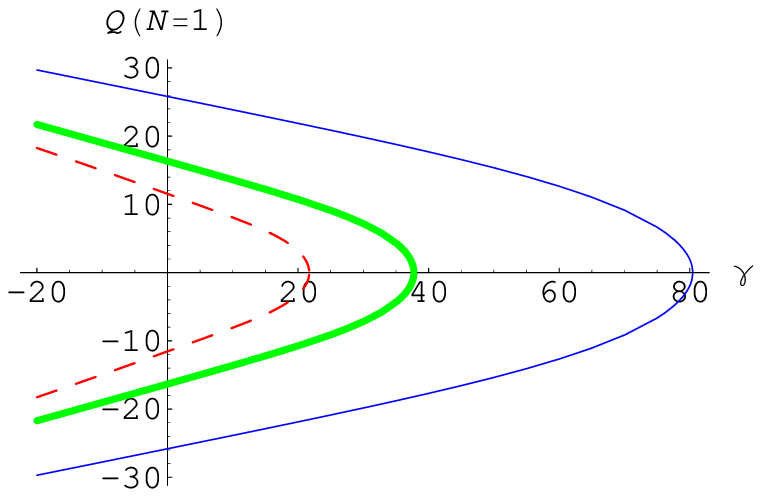}
\includegraphics[width=5.6cm,height=4cm]{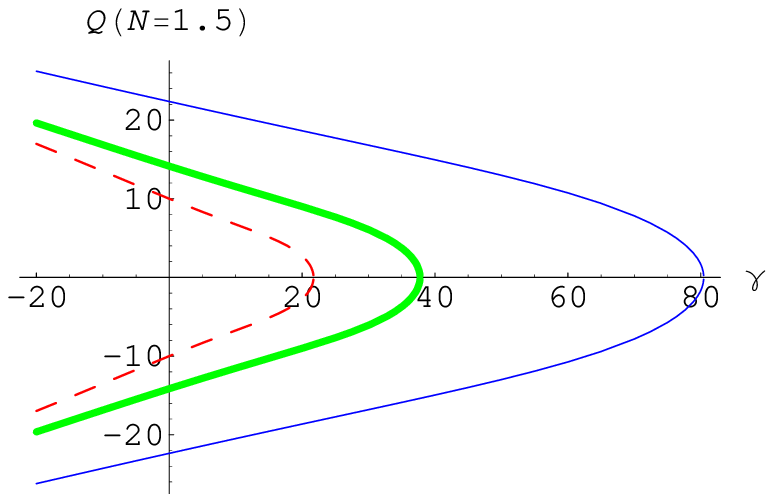}
\end{center}
\vskip -4mm \caption{The value of charge $Q$ at divergence points of
the curvature $\mathcal{R}'_{L}$ vs. parameter $\gamma$. The
parameter $\alpha$ is set to $\frac{\sqrt{3}}{3}$. The entropy is
set to $S=100$ for red dashed lines, $S=200$ for green thick lines
and $S=500$ for blue thin lines.}\label{QVR}
\end{figure*}

Now, we would like to discuss how the geometry behaved as the
parameter $\gamma$ takes different values. For simplicity, we turn
back to the case $N=1$. The Legendre invariant metric for this case
is
\begin{eqnarray}
g_{L}= \left(
  \begin{array}{cc}
    -\frac{B C}{128 \pi^{2}S^{2}(S-\gamma \ln S)^{4}} & 0 \\
    0 & \frac{B}{16 \pi^{2}S^{2}(S-\gamma \ln S)^{2}}  \\
  \end{array}
\right), \label{metric}
\end{eqnarray}
where $B=K|_{N=1}$ and $C=(A_{2}Q^{2}-A_{1})|_{N=1}$
After some tedious calculations, we can obtain the curvature. The
numerator of the curvature is a cumbersome expression and can not be
written in a compact form. While the denominator of it is
proportional to the determinant of the metric (\ref{metric}) and is
given by
\begin{eqnarray}
 D=B^{3}\cdot C^{2}.
\end{eqnarray}
Fixing the parameters $\alpha$ and entropy $S$, the characteristic
behavior of the curvature is depicted in Fig. \ref{2RL}, where the
parameter $\gamma$ is set to $\frac{1}{2}$, $\frac{5}{6}$ and
$\frac{3}{2}$, respectively. The values of the charge $Q$ at the
divergence points of the curvature $\mathcal{R}'_{L}$ are given as
\begin{eqnarray}
 Q=\pm2\sqrt{3}\alpha\sqrt{\frac{{
   (S-\gamma\ln S)\big[\big(1+2\ln{S}\big)\gamma^{2}-4S{\gamma}+S^{2} \big]}}
       {{\big[2(1+\ln S)+1\big]\gamma^{2}-8S\gamma+3S^{2}}}}.
\end{eqnarray}
When $\Delta=3-2\ln S\geq 0$, there are three points of $\gamma$
for the vanished charge $Q$:
\begin{eqnarray}
 \gamma_{1}=\frac{S}{\ln S},\;\;
 \gamma_{\pm}=\frac{2\pm\sqrt{3-2\ln S}}{1+2\ln S}S.
\end{eqnarray}
In general, we consider $S\gg 1$, which leads to $\Delta< 0$. So,
the vanished charge $Q$ is only at $\gamma=\gamma_{1}$. From Fig.
\ref{QVR}, it is easy to conclude that the absolute values of the
charge $Q$ at the divergence points of the curvature
$\mathcal{R}'_{L}$ will decrease with the increase of the
parameter $\gamma$ for large $S$.

\section{Conclusion}
\label{secConclusion}

In this paper, we study the phase transitions and geometry
structure of the plane symmetry black hole. The local
thermodynamic stability of it is also discussed through the heat
capacity $C_{Q}$. It is shown that there always exist locally
thermodynamically stable phases and unstable phases for plane
symmetric black hole due to suitable parameter regimes. The
Weinhold geometry and the Ruppeiner geometry are obtained. The
Weinhold curvature gives phase transition points, which correspond
to that of the first-order phase transition only at $N=1$. While
the Ruppeiner one shows first-order phase transition points for
arbitrary $N\neq 1$. Both of which give no any information about
the second-order phase transition. Quevedo et. al. first pointed
out that the two geometry metrics are not Legendre invariant and
they introduced a Legendre invariant metric, which can give a well
description of various types black hole thermodynamics.
Considering the Legendre invariant, we obtain a unified geometry
metric, which gives a correct behavior of the thermodynamic
interaction and second-order phase transition. Including the
logarithmic corrected term, we study the geometry structure of the
plane symmetry black hole. The result show that the logarithmic
correction term does not affect the unified geometry to depict the
phase transitions of it. We also obtain the result that, for large
$S$, the absolute values of the charge $Q$ at the divergence
points of the curvature $\mathcal{R}'_{L}$ will decrease with the
increase of the parameter $\gamma$. In this paper, we show that
the unified geometry description gives a well description of the
second-order phase transitions of the plane symmetry black hole.
We also expect that this unified geometry description may give
more information about a thermodynamic system.

\section*{Acknowledgements}
This work was supported by the Program for New Century Excellent
Talents in University, the National Natural Science Foundation of
China (No. 10705013), the Doctoral Program Foundation of
Institutions of Higher Education of China (No. 20070730055), the
Natural Science Foundation of Gansu province, China (No.
096RJZA055), the Key Project of Chinese Ministry of Education (No.
109153), and the Fundamental Research Fund for Physics and
Mathematics of Lanzhou University.

\end{document}